\documentclass[twocolumn,aps]{revtex4}

\usepackage{graphicx}
\usepackage{amsmath}
\usepackage{amssymb}
\usepackage{enumerate}
\usepackage{tabularx}
\usepackage{siunitx}
\usepackage{xcolor}
\usepackage{braket}
\usepackage{mathtools}
\usepackage{tikz}
\usetikzlibrary{matrix,decorations.pathreplacing,calc}

\pgfkeys{tikz/mymatrixenv/.style={decoration=brace,every left delimiter/.style={xshift=0pt},every right delimiter/.style={xshift=-0pt}}}
\pgfkeys{tikz/mymatrix/.style={matrix of math nodes,left delimiter=(,right delimiter={)},inner sep=2pt,column sep=1em,row sep=0.5em,nodes={inner sep=0pt}}}
\pgfkeys{tikz/mymatrixbrace/.style={decorate,thick}}

\newcommand\mymatrixbraceoffsetv{0.5em}

\newcommand*\mymatrixbracebottom[4][m]{
    \draw[mymatrixbrace] ($(#1.south west)!(#1-1-#3.south east)!(#1.south east)-(0,\mymatrixbraceoffsetv)$)
        -- node[below=2pt] {#4} 
        ($(#1.south west)!(#1-1-#2.south west)!(#1.south east)-(0,\mymatrixbraceoffsetv)$);
}

\begin{document}

\title{Trapped ion quantum error-correcting protocols using only global operations}

\author{Joseph F. Goodwin}
\email{joseph.goodwin09@imperial.ac.uk}
\affiliation{Quantum Optics and Laser Science, Blackett Laboratory, Imperial College London, Prince Consort Road, London, SW7 2AZ, United Kingdom.}

\author{Benjamin J. Brown}
\affiliation{Quantum Optics and Laser Science, Blackett Laboratory, Imperial College London, Prince Consort Road, London, SW7 2AZ, United Kingdom.}

\author{Graham Stutter}
\affiliation{Quantum Optics and Laser Science, Blackett Laboratory, Imperial College London, Prince Consort Road, London, SW7 2AZ, United Kingdom.}

\author{Howard Dale}
\affiliation{Quantum Optics and Laser Science, Blackett Laboratory, Imperial College London, Prince Consort Road, London, SW7 2AZ, United Kingdom.}

\author{Richard C. Thompson}
\affiliation{Quantum Optics and Laser Science, Blackett Laboratory, Imperial College London, Prince Consort Road, London, SW7 2AZ, United Kingdom.}

\author{Terry Rudolph}
\affiliation{Quantum Optics and Laser Science, Blackett Laboratory, Imperial College London, Prince Consort Road, London, SW7 2AZ, United Kingdom.}

\begin{abstract}
Quantum error-correcting codes are many-body entangled states used to robustly store coherent quantum states over long periods of time in the presence of noise. Practical implementations will require efficient entangling protocols that minimise the introduction of noise during encoding or readout. We propose an experiment that uses only global operations to encode information to either the five-qubit repetition code or the five-qubit code on a two-dimensional ion Coulomb crystal architecture. We show we can prepare, read out, and acquire syndrome information for these two codes using only six and ten global entangling pulses respectively. We provide an error analysis, estimating we can achieve a six-fold improvement in coherence time with as much as 1\% noise in the control parameters for each entangling operation.
\end{abstract}

\maketitle

\section{Introduction}
The endeavour towards large-scale quantum technologies has recently seen impressive experimental efforts in the realisation of small quantum error-correcting codes (QECCs) using ion traps~\cite{Chiaverini,Nigg, Lanyon}, and also superconductors~\cite{Gladchenko, Reed, Barends, Kelly15, Corcoles15}. Given sufficient control over the composite physical systems, coherence times of encoded quantum states will increase exponentially with the size of the code. Ultimately we hope to achieve the experimental prowess required to build arbitrarily large codes, but a modest intermediate goal is the development of small error-correcting codes that can maintain quantum coherence for time scales longer than those of their composite parts~\cite{Tomita14}. Such codes will be useful for building technologies using distributed quantum architectures~\cite{Nickerson13, Monroe, Nickerson14}, and provide long coherence times capable of interrogating fundamental aspects of quantum mechanics~\cite{Simon03}.

\begin{figure}
\includegraphics[width=\columnwidth]{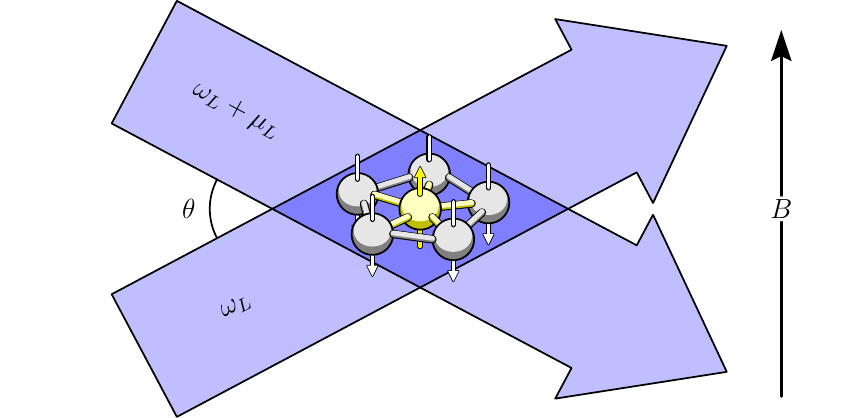}
\caption{Quantum error-correcting codes are prepared using six ions trapped in a rotationally symmetric planar crystal. The trap magnetic field is $B$. A periodic spin-dependent force is generated by a pair of off-resonant laser beams with frequency difference $\mu_L$, and angular separation of $\theta$, symmetric above and below the crystal plane. We encode and readout information to the crystal using only global operations by variation of $\mu_L$, along with a standard set of collective local microwave rotations.\label{ExperimentalSetup} }
\end{figure}

The difficulties in the realisation of QECCs in ion traps are technical, first among them the experimental finesse necessary to perform two-ion entangling gates on an array of many ions, as any spectator ions must be decoupled from the entangling interaction. This can be achieved physically, via shuttling~\cite{Chiaverini}, or spectroscopically, either by `hiding' ions in ancillary states~\cite{Nigg}, or by using dynamic decoupling techniques~\cite{Lanyon}. Both spectroscopic methods require repeated addressing of individual ions.

In this Paper we propose an alternative to the local circuit approach for realising many-body entangled states. We describe an experimental protocol to prepare two QECCs using global entangling operations alone, avoiding the need to decouple subsets of the qubit register. 

We consider a six-ion Coulomb crystal in a Penning trap~\cite{Mavadia} as depicted in Fig.~\ref{ExperimentalSetup}, and we provide a protocol to encode an arbitrary quantum state to either the five-qubit code (5QC)~\cite{DiVincenzo}, or the five-qubit repetition code (5RC). We extract syndrome information while teleporting information from the code, allowing us to correct for errors that may have occurred while the logical state was stored. We provide a noise analysis showing that we double coherence times with an entangling pulse noise of 1.7\% and 0.2\% for 5RC and 5QC respectively. Such performance is attainable using current technology.

In general, it is difficult to find global unitary operations that realise a chosen target state from a given input state. However, many target states share a common symmetry with the crystal architecture. This simplifies protocols to realise desired states, as all the terms in the state that are equivalent up to the symmetry evolve identically. The global pulse sequences we discover enable us to execute entire error-correcting protocols, including syndrome extraction, using a very low number of operations compared with other trapped-ion protocols~\cite{Lanyon, Chiaverini, Nigg}. We can perform 5RC(5QC) using only 14(18) discrete operations. The number of operations required to perform these protocols with two-qubit entangling gates is typically an order of magnitude greater. Conventional non-demolition stabilizer measurements are even more taxing; the 5QC stabilizers require several hundred operations per cycle. In this respect our protocol is favourable, as it substantially reduces the total gate noise introduced to the system~\cite{Knill, Steane}. 

This paper is structured as follows. In Sec.~\ref{SEC:ExperimentalSetup} we introduce the general architecture and describe our approach to generating symmetry-preserving global unitaries. Sec.~\ref{SEC:QECProtocols} outlines how these techniques can be applied to a six-ion Coulomb crystal to perform two quantum error-correction protocols. In Sec.~\ref{SEC:Search} we describe the numerical methods used to find pulse sequences to engineer the necessary unitaries. We subsequently provide a noise analysis of the proposed experiments in Sec.~\ref{SEC:NoiseAnalysis}. We give some concluding remarks in Sec.~\ref{SEC:Conclusion}. Technical details of our experimental proposals are given in Apps.~\ref{APP:FlorescencePatterns} and~\ref{APP:PulseSequence}.

\section{Global Unitary Operations}
\label{SEC:ExperimentalSetup}

\begin{figure}
\includegraphics{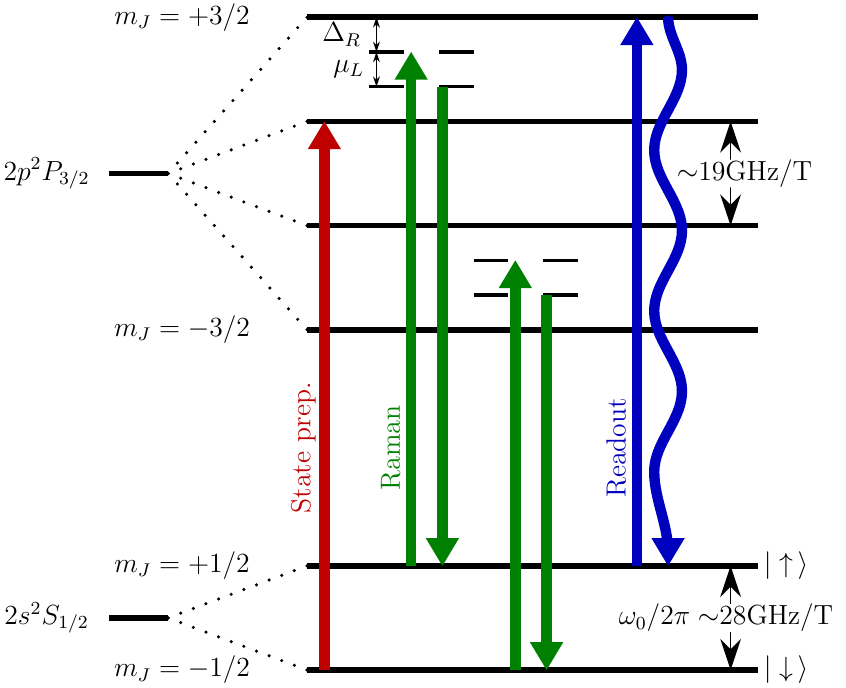}
\caption{Schematic of optical dipole force and readout lasers for $^9\text{Be}^+$. At typical magnetic field strengths, the detuning of the Raman lasers from the $S_{1/2}\rightarrow P_{3/2}$ transition, $\Delta_R$, is several tens of GHz, while the difference frequency between the upper and lower beams is $\mu_L\approx1$MHz.}\label{Transitions}
\end{figure}

We consider a qubit register in a two-dimensional ion Coulomb crystal. Physical qubits are written to the two Zeeman sublevels of the $S_{1/2}$ ground state of each ion. The protocols we describe rely on the Zeeman splitting being large and are thus best suited for implementation in a Penning trap, where these levels are separated by $\sim\SI{100}{\giga\hertz}$ at typical magnetic field strengths. The state of the qubit register of $N$ ions is written in the spin basis $\ket{s}_R \equiv \ket{s_1}\otimes \ket{s_2}\otimes \dots \otimes \ket{s_N}$ for $s_j = \uparrow,\downarrow$, where arbitrary states are written 
\begin{equation}
 \ket{\chi}_R=\sum_{s} r_s\ket{s}_R ,
\end{equation}
 such that $\sum_s|r_s|^2 = 1$. We do not consider a particular species of ion, but suitable examples include magnesium and beryllium, which we show in Fig.~\ref{Transitions}.

Projective readout of the physical qubits in the computational basis is given by fluorescence from a laser resonant with a suitable dipole transition, for instance $S_{1/2}\rightarrow P_{3/2}$. This transition is also used for Doppler cooling the crystal. A microwave field at the qubit frequency $\omega_0$ permits global Pauli $X$ and $Y$ rotations, while Pauli $Z$ rotations are effected by applying the laser normally used for readout, detuned far from resonance, producing an AC Stark shift of the qubit levels.

We exploit the coupling between transverse mechanical modes of the Coulomb crystal and spin-dependent optical dipole forces (ODFs) to realise complex unitary operations on the qubit register. We show that we can generate global unitary operations of the form
\begin{equation}
U(\boldsymbol\phi)=\sum_{s}e^{i \phi_s}\ket{s}\bra{s}_R, \label{GlobalUnitary}
\end{equation}
where $\boldsymbol\phi$ is a vector of phases $\phi_s$ determined by the spin-mode couplings.

Following the treatment of Britton {\em et al}~\cite{Britton2012}, we generate a spin-dependent ODF by application of a pair of crossed Raman lasers with detuning $\Delta_R\sim\SI{40}{\giga\hertz}$ 
from a suitable dipole transition, shown in Fig.~\ref{Transitions}. The polarisations of the two beams can be adjusted such that the AC Stark shift of the qubit transition frequency $\omega_0$ due to each individual beam is zero, while their interference produces a one-dimensional optical lattice with wavevector $\textbf{k}_L=(0,0,k_L)$, the polarisation gradient of which provides the force. Additionally, we introduce a frequency difference $\mu_L$ between the lasers causing the optical lattice to scan across the crystal and leading to a periodic, transverse driving force. If this frequency difference is tuned close to one of the normal mechanical modes of the crystal, the ODF can be made to excite collective vibrations of the ions.

In general, a quantum harmonic oscillator driven off-resonantly will traverse a closed loop in phase space with radius proportional to the driving force~\cite{Ozeri}. The system returns to its initial motional state at times $\tau=2\pi/\delta$ where $\delta$ is the detuning from resonance, acquiring a geometric phase proportional to the area of the loop. If the driving force is spin-dependent then this can be used to entangle the spin degree of freedom, as in the two-ion phase gate~\cite{Milburn, Leibfried}. We demonstrate the extension of this technique to arbitrary numbers of ions driven in modes that do not couple identically to all ions. Unlike Ref.~\cite{Britton2012} and other previous applications of such ODF beams, we do not require the beams to produce forces of equal magnitude on opposing spins, allowing their ratio $R=F_\uparrow/F_\downarrow$ to be varied between approximately -0.5 and -2 through adjustment of $\Delta_R$. This modification greatly increases the range of unitaries that may be produced.

We obtain the transverse modes of the Coulomb crystal with the methods described in~\cite{Freericks}, solving the eigenvector equation for the $N\times N$ axial stiffness matrix $\textbf{K}^{zz}$. The mode eigenvectors $\ket{\textbf{a}_m}$ are the eigenvectors of $\textbf{K}^{zz}$ and the corresponding mode eigenfrequencies $\omega_{m}$ are related to its eigenvalues $\lambda$ by the mass of a single ion, $M$, as $\omega_m=\sqrt{\lambda_{m}/M}$. 

Writing the set of mode eigenvectors as a matrix $\textbf{A}=\sum_i \ket{i}\bra{\textbf{a}_m}$ with elements $A_{mi}$, we can decompose any forces $f_i$ on the constituent ions into generalised forces $Q_m$ acting on each mode, where
\begin{equation}
Q_m=\sum_i A_{mi}f_i.
\end{equation}

Given a uniform ODF beam across the crystal, we express the forces on the whole register as a matrix $\textbf{F}$, with elements $F_{is}$ representing the forces on ion $i$ for each $\ket{s}_R$ in $\ket{\chi}_R$. The product of the modal matrix $\textbf{A}$ and this new dipole force matrix gives the spin-mode coupling matrix $\textbf{M}$ with elements
\begin{equation}
M_{ms}=\sum_i A_{mi}F_{is}.
\end{equation}
The matrix $\textbf{M}$ defines the generalised force acting on each normal mode for each state $\ket{s}_R$ of the register. In the case of a symmetric crystal, many of the modes will have a degenerate partner of the same frequency. A pair of degenerate modes will be driven simultaneously, so it is proper to consider the Pythagorean sum of the generalised forces on each pair and the corresponding rows in $\textbf{M}$ are combined in quadrature.

We apply a set of pulses near-resonant with each mode such that $\mu_L=\omega_m-\delta$, where we define a vector $\mathbf{P}$ with elements $P_m$ proportional to pulse areas, and we use the freedom to adjust $\delta$ to ensure the loops close. The geometric phase acquired by each basis states in $\ket{\chi}_R$ is proportional to the square of the generalised force, and given by
\begin{equation}
\phi_s=\sum_m(M_{ms})^2 P_m,
\end{equation}
determining the unitary of Eqn.~(\ref{GlobalUnitary}).

\section{Error-Correction Protocols}
\label{SEC:QECProtocols}

\begin{figure}
\includegraphics{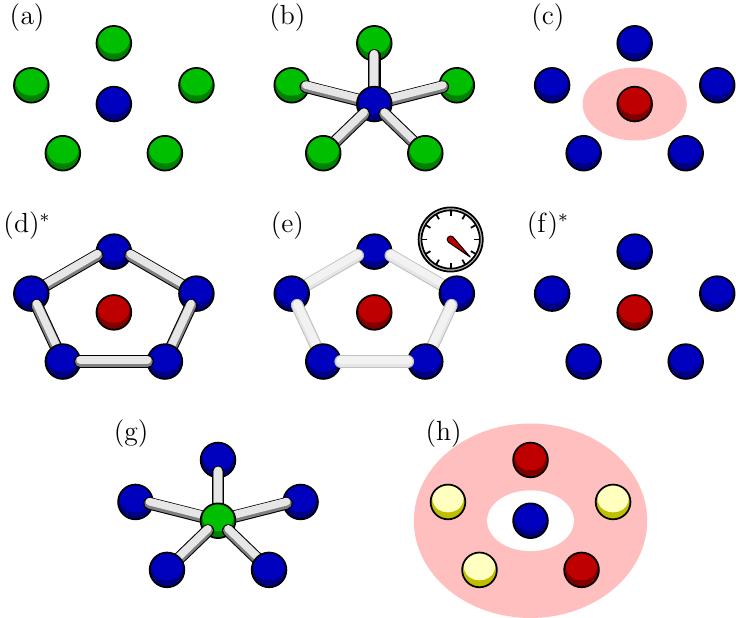}
\caption{The protocol for 5RC and 5QC. Figures marked with an asterisk are omitted for 5RC. 
(a) We initialise the crystal in the product state, $\ket{\chi_0}_R $. The hub qubit, shown in blue, lies at the centre of the crystal. The code qubits of the crystal, shown in green, form a rotationally symmetric outer ring. 
(b) We perform unitary $U_{\text{spokes}}$. This entangles the hub qubit to the 5RC codespace. (c) We measure the central ion to teleport the logical information onto the blue ring ions, completing the encoding of 5RC. (d) (Omitted for 5RC) To prepare 5QC, we also apply $U_{\text{ring}}$ immediately after the measurement, mapping codewords of 5RC to 5QC. (e) The logical qubit is protected now the code is prepared. (f) (Omitted for 5RC) To begin readout for 5QC, we apply $U_{\text{ring}}$ to map 5QC codewords to 5RC codewords. (g) We prepare the central qubit in the $\ket{+} = (\ket{\uparrow} + \ket{\downarrow}) / \sqrt{2}$ state and perform $U_{\text{spokes}}$ once more, entangling the codespace to the hub. (h) Measuring the code qubits teleports the quantum information back to the central ion. As the Pauli-$X$ measurements we perform commute with the stabilizers of 5RC, the fluorescence pattern of the measured bright and dark ions, shown in red and yellow, provides syndrome information (see Appendix~\ref{APP:FlorescencePatterns}). \label{Protocol}}
\end{figure}

We now show we can use the described experimental setup to prepare and read-out the five-qubit code (5QC), and the five-qubit repetition code (5RC), in a six-ion register. The 5QC and 5RC encode a single logical qubit in the subspace of the Hilbert space of five physical qubits. Additionally, we measure syndrome information that enables us to identify errors to the encoded state, which we can then correct with an appropriate single qubit operation. The 5QC protects the encoded information against arbitrary Bloch sphere rotations suffered by a single physical qubit of the register. Alternatively, the 5RC protects against dephasing channel errors acting on up to two physical qubits, but cannot correct for spin-flip noise. In many ion qubits, the $\text{T}_1$ lifetime is several hundred seconds. In these cases, the 5RC provides sufficient protection against a realistic noise model.

We consider the explicit example of a qubit register of six ions in a crystal in the configuration shown in Fig.~\ref{Protocol}(a). The qubit of the central `hub' ion is initialised in the logical state 
\begin{equation}
\ket{\psi}_H = \alpha \ket{\downarrow}_H + \beta \ket{\uparrow}_H.
\end{equation} 
The five `code' ions in the ring around the hub will encode the quantum information. They are initially prepared in the product state
\begin{equation}
\ket{S}_C = | + + + + + \rangle_C = \frac{1}{2^{5/2}}\sum_s \ket{s}_C, 
\end{equation} 
where the summation is made over all configurations in the spin basis of the five code ions. Given the ability to prepare the initial state 
\begin{equation}
\ket{\chi_0}_R = \ket{\psi}_H \otimes \ket{S}_C, \label{Eqn:InitialState}
\end{equation} 
and the ability to measure qubits of the register in the Pauli-$X$ basis, we require only two global unitary operations, $U_{\text{ring}}$ and $U_{\text{spokes}}$, to encode and read out the 5QC, and only $U_{\text{spokes}}$ to encode and read out 5RC. The unitaries $U_{\text{spokes}}(U_{\text{ring}})$ will globally apply a controlled phase gate between each of the qubits in the register that share a grey edge in Fig.~\ref{Protocol}(b)(\ref{Protocol}(d)). We summarise the error-correcting protocol in the caption of Fig.~\ref{Protocol}. 

The protocol we described requires the ability to prepare and measure the hub qubit independently of the code qubits and vice versa. This can be achieved with a focussed beam, or globally, using two isotopes of a suitable ion species. Here the isotopic shift would be sufficient to allow resonant frequency addressing during state preparation and readout, without significantly affecting the coupling to the off-resonant entangling pulses. We also require the ability to globally realise unitary operations $U_{\text{ring}}$ and $U_{\text{spokes}}$. Finding pulse sequences to realise such global operations can be challenging. In the following Section we describe the numerical methods we use to find suitable sequences.

\section{Searching for Global Unitary Operations}
\label{SEC:Search}

Given suitable control to prepare the initial state and the ability to make the necessary measurements, we must also be able to perform the unitary operations $U_{\text{ring}}$ and $U_{\text{spokes}}$ to complete the described error-correction tasks. To do so, we need to find a pulse sequence that simultaneously achieves phases $\phi_s$  that correspond to the desired unitary operation. In this section we explain the numerical methods we use to find pulse sequences that realise the required rotations.

A six-ion planar crystal has six transverse modes including two degenerate pairs, shown in the matrix $\textbf{A}^T$ in order of decreasing frequency
\begin{align*}
\mathbf{A}^T=
\begin{tikzpicture}[mymatrixenv, baseline=-0.5ex]
   \matrix[mymatrix,ampersand replacement=\&] (m) {
1 \& 0 \& 0 \& 0 \& 0 \& -5 \\
1 \& (G-2) \& 0 \& -(G+1) \& (1-G) \& 1 \\
1 \& 2(1-G) \& G \& 2G \& 0 \& 1 \\
1 \& (G-2) \& -G \& -(G+1) \& (G-1) \& 1 \\
1 \& 1 \& -1 \& 1 \& -1 \& 1 \\
1 \& 1 \& 1 \& 1 \& 1 \& 1 \\
    };
    \mymatrixbracebottom{1}{1}{$\omega_1$}
\mymatrixbracebottom{2}{3}{$\omega_2$}
\mymatrixbracebottom{4}{5}{$\omega_3$}
\mymatrixbracebottom{6}{6}{$\omega_4$}
\end{tikzpicture}
\end{align*}
where $G = (1+\sqrt{5})/2$ and the mode vectors are shown unnormalised for brevity. We have five control parameters with which to engineer the desired unitary, the pulse areas, $P_m$, applied to each of the four mode frequencies, and the force ratio, $R$.

In general it is unlikely that we can find a pulse sequence that simultaneously generates an arbitrary choice of target phases, $\phi_s'$. However, we are able to significantly increase the search space by exploiting our freedom to choose an arbitrary global phase, $\phi_g$, and by taking $\phi_s$ values {\em modulo $2\pi$}. We define the {\em total phase}, $\Phi_s(n_s)$ to be
\begin{equation}
\Phi_s(n_s) = \phi_s + \phi_g + 2\pi n_s,
\end{equation}
where $n_s$ takes integer values. Finding a pulse sequence that leads to phases $\Phi_s(n_s)$ instead of $\phi_s$ will achieve the same effective unitary evolution.

The problem thus consists of a series of conventional non-linear optimisation repeated across a $\mathcal{O}(2^N)$-dimensional space of integer strings, $\mathbf{n}$.
Within the continuous parameter space of each integer string, we use a Nelder-Mead search to minimise a cost function
\begin{equation}
C(\mathbf{P},R, \mathbf{n}) = \sum_s  \left[ \Phi_{s+1}- \phi'_{s+1} -\Phi_s+\phi'_s \right]^2 = 0.\label{Eqn:CostFunction}
\end{equation}
The difficulty of the optimisation problem lies in searching the space of integer strings. To simplify the problem we bound the integer values within the limits $-10\le n_s\le10$. This restriction is experimentally advantageous; small values of $n_s$ are favourable as large phase space areas are more sensitive to errors in pulse length or intensity. With this restriction, the systems we describe are small enough that employing exhaustive search techniques is not unreasonable. We make use of the fact that each element in the summation on the right-hand side of Eqn.~(\ref{Eqn:CostFunction}) is non-negative, and therefore the terms of the series must be zero for a solution to hold. We are then able to perform a series of simpler minimisation problems for shorter strings of integers to find integer solutions for only a few terms of the cost function at once. Completing these simpler search problems enables us to efficiently eliminate large regions of the integer search space, and thus significantly speeds up the exhaustive search. 

Using the described techniques, we find a large number of sparsely distributed solutions for our target unitaries; we are free to select those that will minimise the sensitivity to expected experimental noise. As well as minimising the total pulse areas, it is useful to select solutions with modest force ratios, $R$, as these are less sensitive to polarisation noise in the optical dipole force lasers. One solution for $U_{\text{spokes}}$ is 
\begin{equation}
(P_1,P_2,P_3,P_4,R) = (3.125, 2.604, 2.604, 0,-1.400), \label{SpokePulseSequence}
\end{equation} 
where the pulse areas are normalised such that a centre-of-mass pulse $P_1=1$ would produce a $\pi$ radian phase shift on the $\ket{\downarrow}_H \otimes \ket{\downarrow \downarrow \downarrow \downarrow \downarrow}_C$ state.  We give a full description of the state evolution during this pulse sequence in App.~\ref{APP:PulseSequence}. The values of the integer string for the minimisation we find takes small values, with $-3 \le n_s \le 3$ for all $s$.

Additionally, we find the pulse squence 
\begin{equation}
(P_1,P_2,P_3,P_4,R) = (10.99, 7.677, 19.65, 10.99, -0.6737),
\end{equation} 
will produce $U_{\text{ring}}$. We remark that we can vary the ratio of forces for each pulse, but for the present purposes it is sufficient to keep this constant throughout the sequence. 

Another possible solution to the integer search would be the use of evolutionary algorithms, which would not provide an exhaustive search but might be more efficient for large numbers of qubits. We have found that simple evolutionary algorithms enable us to find some of the solutions that we obtain using an exhaustive approach.

While numerical search methods work for the codes we describe here, and may be used to obtain other interesting unitary operations on small registers, the size of the integer space suggests that this approach will not scale well to larger registers. Finding solutions for much larger problems will likely depend on the whole or partial use of analytical methods for determining suitable pulse sequences. This is the subject of ongoing work~\cite{Seis14}.

\section{Noise Analysis}
\label{SEC:NoiseAnalysis}

Our results show that we can execute both QECC protocols using very few pulses. Short pulse sequences are advantageous as errors that occur during the protocol are minimised. We numerically simulate both protocols using imperfect unitary operations to estimate the engineering requirements necessary to reduce the decay of quantum coherence.

\begin{figure}
\includegraphics[width=\columnwidth]{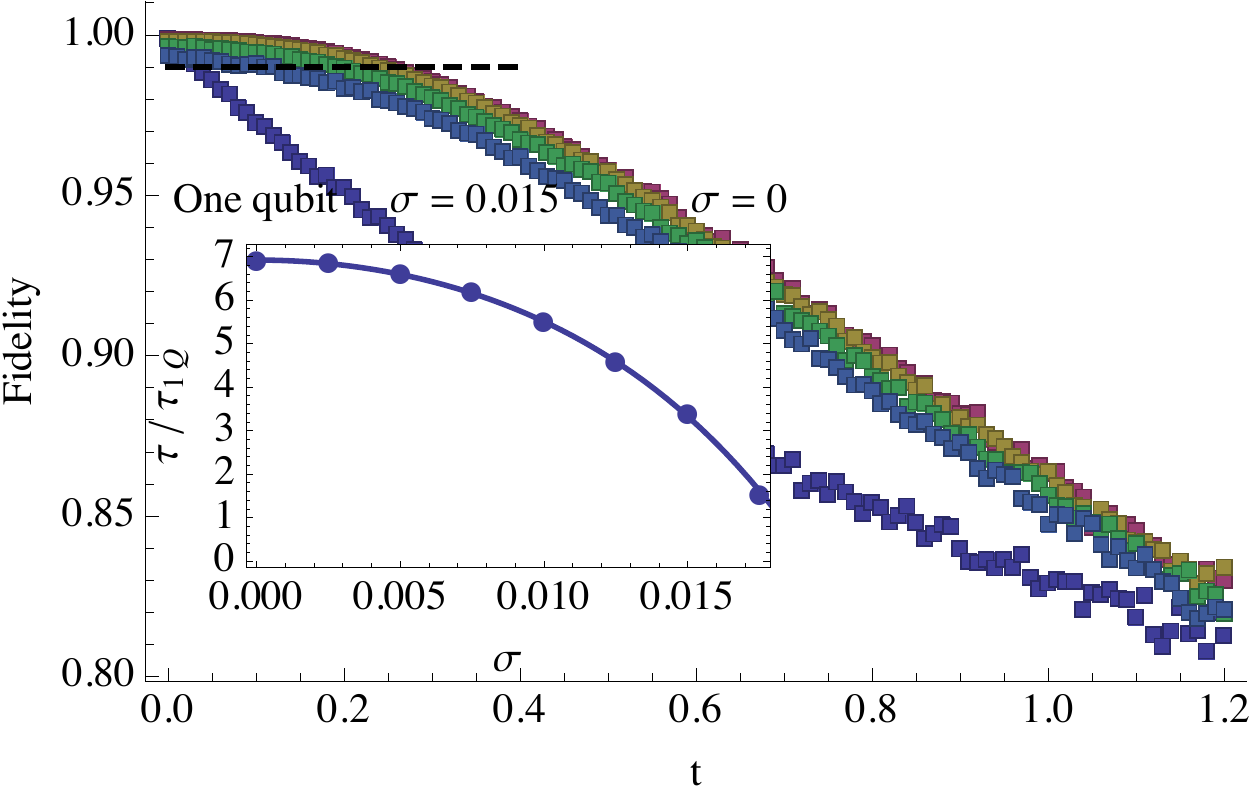}
\caption{\label{Fidelities}Average fidelity as a function of time for imperfectly prepared 5RC states, with time measured in units of ion coherence times. Protocol executed with pulse noise $\sigma = 0,\,0.005,\,0.01$ and $0.015$ shown (from top to bottom) in red, yellow, green and blue respectively. The fidelity of a single qubit under the dephasing map is also shown in blue (lower line). Dashed line marks 0.99 fidelity. The inset shows unitless ratio $\tau / \tau_{\text{1Q}}$ plotted as a function of the standard deviation of pulse noise $\sigma$. A fit to the data yields $\tau/\tau_{1Q} \sim \tau_{0}(2 - \exp( 1858 \sigma^2 ))$.}
\end{figure}

 We examine the decoherence of the encoded state over time $t$ where we perform the protocol using imperfect unitary operations. We numerically simulate the 5RC protocol using the imperfect $U_{\text{spokes}}^{\sigma}$ unitary, where we disrupt the optimal pulse sequence for $U_{\text{spokes}}$, Eqn.~(\ref{SpokePulseSequence}), by replacing pulse areas $P_m$ and force ratio $R$ with $\epsilon_m(\sigma) P_m$ and $\epsilon_R(\sigma)R$ respectively, where $\epsilon_m(\sigma)$ and $\epsilon_R(\sigma)$ are random variables chosen from a normal distribution with standard deviation $\sigma$ and unity mean. The spin of each individual ion, $\rho$, dephases via the completely positive and trace preserving map 
 \begin{equation}
 \gamma_t(\rho) = (1+e^{-t})\rho /2 + (1-e^{-t}) Z \rho Z / 2, \label{Eqn:DephasingNoise}
 \end{equation} 
 where $Z$ is the Pauli-Z matrix. Importantly, $\gamma_t$ is continuous in $t$, and converges to the maximally dephased state in the $t \rightarrow \infty $ limit, i.e.
 $$ \lim_{t\rightarrow \infty}\gamma_t(\rho)  = ( \rho +  Z \rho Z ) / 2 . $$
 
We additionally assume that each application of $U^\sigma_{\text{spokes}}$ is performed in time $t_{U} = 5\cdot10^{-4}$, equivalent to a 50$\mu$s gate for qubits with 100ms coherence time. We model the noise introduced during the application of $U^\sigma_{\text{spokes}}$ by applying the dephasing map to each qubit for a time $t_{U}$ before the instantaneous application of $U^\sigma_{\text{spokes}}$.

We measure decoherence using the average fidelity 
\begin{equation}
\mathcal{F}(\rho_H) = \left| \text{tr}[\rho_H \ket{\psi}\bra{\psi}_H] \right|, 
\end{equation} 
where $\rho_H$ is the final state that is read out from the code, to be compared with the initially encoded state, $\ket{\psi} \bra{\psi}_H$. We are interested in the decay of fidelity over time. In Fig.~\ref{Fidelities} we plot $\mathcal{F}(\rho_H)$ as a function of time. The fidelity is calculated with $500$ random samples of $\ket{\psi}_H$ and $U^\sigma_{\text{spokes}}$ for each $t$. 

To quantify the improvement in coherence time attained using the code, we compare its coherence time to that of a qubit stored in the spin of a single ion. We define the high-fidelity time, $\tau$, to be the time the prepared code maintains fidelity above 0.99. We compare $\tau$ to $\tau_{1\text{Q}}$, the high-fidelity time of a single qubit. The inset of Fig.~\ref{Fidelities} shows $\tau / \tau_{1\text{Q}}$ for 5RC as a function of $\sigma$. We observe scaling
\begin{equation}
\tau / \tau_{1\text{Q}}  = \tau_0(2 - e^{ \sigma^2 / \alpha}), 
\end{equation} 
where $\tau_0 = \tau/ \tau_{1\text{Q}}$ is the code-dependent optimal coherence time improvement we can achieve for the noiseless $\sigma = 0$ case. Our results show that we converge exponentially towards $\tau_0$ as $\sigma^2$ decreases. For 5RC, we find $\alpha \sim 5.4\cdot 10^{-4}$, and $\tau_0 \sim 6.92$. 

We perform a similar simulation for 5QC, using disrupted $U_{\text{ring}}$ and $U_{\text{spokes}}$ unitaries, 
where ions decohere via the depolarising map 
\begin{equation}
\xi_t(\rho) = (1 + 3e^{-t})\rho / 4 + (1-e^{-t})(X\rho X + Y \rho Y + Z \rho Z) / 4. \label{Eqn:DepolarisingNoise}
\end{equation} 
Like the dephasing map given in Eqn.~(\ref{Eqn:DephasingNoise}), the map~$(\ref{Eqn:DepolarisingNoise})$ is continuous in the parameter $t$, and in the long time limit the map converges to the maximally depolarised state, namely
\begin{equation}
 \lim_{t\rightarrow \infty} \xi_t (\rho)= ( \rho + X \rho X + Y\rho Y + Z\rho Z ) / 4. 
 \end{equation}
Using the depolarising noise map and $50\mu$s pulse times, we find $\alpha \sim 3.0 \cdot 10^{-5}$ and $\tau_0 \sim 2.45$ for the 5QC protocol. 

We extrapolate the present data to the value $\tau / \tau_{1\text{Q}} = 1$ to obtain a threshold pulse noise of $\sigma_{\text{th}}\sim 0.0038$ for 5QC, and $\sigma_{\text{th}}\sim 0.018$ for 5RC. Producing gates that meet these threshold values is feasible with existing technology. We expect further improvements to be achieved by repeating the protocol periodically, thus suppressing errors before significant decoherence has occurred.

{\color{blue}

}

\section{Conclusion}
\label{SEC:Conclusion}

We have shown, by utilising symmetries of the mechanical modes of an ion Coulomb crystal, that we can find global entangling operations to produce interesting quantum states. We have demonstrated this by describing two quantum error-correcting protocols using the presented scheme. Further, we have shown that the simplicity of the proposed QECC protocols leads to improvements in quantum coherence with relatively modest experimental demands. We expect the present experimental proposal will produce results that are competitive with current state of the art trapped-ion error-correcting protocols~\cite{Chiaverini,Nigg, Lanyon}. One could use the protected qubits of the present proposal in a hybrid distributed scheme to realize scalable computation~\cite{Nickerson13, Monroe, Nickerson14}. In such a scheme, the improvements in coherence times provided by our proposal can lead to exponential improvements in the resource costs of an error-correction code~\cite{Terhal13}. We hope the methods we have used will motivate further study into other entangled states that can be produced entirely or in part by simple global operations.

\begin{acknowledgements}
We thank D. Segal for critical reading of the manuscript. The authors are supported by the EPSRC.\\ 
\end{acknowledgements}

\bibstyle{plain}

\appendix

\section{Syndrome Readout from Florescence Patterns}
\label{APP:FlorescencePatterns}

\begin{table*}[b]
\begin{tabular}{ccc}
\hline\hline
\raisebox{12pt}

Projected Error on 5RC  & Fluorescence Pattern & Logical Pauli Correction\\
\hline\hline
\vspace{-0.6em}\\

$IIIII$&$\Box\Box\Box\Box\Box$&$I$\\
$IIIII$&$\blacksquare\blacksquare\blacksquare\blacksquare\blacksquare$&$X$\\
$IIZII$&$\Box\Box\blacksquare\Box\Box$&$I$\\
$IIZII$&$\blacksquare\blacksquare\Box\blacksquare\blacksquare$&X\\
$ZIIIZ$&$\blacksquare\Box\Box\Box\blacksquare$&$I$\\
$ZIIIZ$&$\Box\blacksquare\blacksquare\blacksquare\Box$&$X$\\
$IZIZI$&$\Box\blacksquare\Box\blacksquare\Box$&$I$\\
$IZIZI$&$\blacksquare\Box\blacksquare\Box\blacksquare$&$X$\\

\hline\hline
\raisebox{12pt}

Projected Error on 5QC & Fluorescence Pattern & Logical Pauli Correction\\
\hline\hline
\vspace{-0.6em}\\
$IIIII$&$\Box\Box\Box\Box\Box$&$I$\\
$IIIII$&$\blacksquare\blacksquare\blacksquare\blacksquare\blacksquare$&$X$\\
$IIZII$&$\Box\Box\blacksquare\Box\Box$&$I$\\
$IIZII$&$\blacksquare\blacksquare\Box\blacksquare\blacksquare$&X\\
$IIYII$&$\blacksquare\Box\Box\Box\blacksquare$&$-Y$\\
$IIYII$&$\Box\blacksquare\blacksquare\blacksquare\Box$&$iZ$\\
$IIXII$&$\Box\blacksquare\Box\blacksquare\Box$&$Z$\\
$IIXII$&$\blacksquare\Box\blacksquare\Box\blacksquare$&$iY$\\
\hline\hline
\end{tabular}
\caption{Projected errors, syndromes and the corresponding Pauli corrections to the final teleported state for 5RC and 5QC. Syndromes are given as florescence patterns of bright and dark ions around the ring during the final readout stage. We represent florescence patterns of dark and bright ions with symbols $\blacksquare$ and $\Box$, respectively. The correction we apply is invariant under cyclic permutations of the florescence pattern.}
\label{tab:syndromes}
\end{table*}

Table~\ref{tab:syndromes} shows a list of correctable Pauli errors, the florescence patterns we expect to observe at readout for a given error, and the corresponding Pauli corrections we must apply when the state is read out.

\section{State Evolution during `Spoke' Pulse Sequence}
\label{APP:PulseSequence}
Table~\ref{table:phases} shows how each basis state evolves under the application of the series of pulses for $U_{\text{spokes}}$, from the initial product state, Eqn.~(\ref{Eqn:InitialState}), in the leftmost column, towards the target state in the rightmost column. The columns list the absolute phases and phase angles after each pulse. We point out that cyclic permutations of ring qubit configuration all acquire the same phase. The Table therefore only shows the phase for one representative state for each configuration that is equivalent up to a cyclic permutation. The bottom row lists the fidelity of each intermediate state relative to the target.

\begin{table*}[b]

\begin{tabularx}{\textwidth}{cc>{\centering}X>{\centering}Xc>{\centering}X>{\centering}Xc>{\centering}X>{\centering}Xc}
    \hline\hline
    \raisebox{12pt}

    Basis $\ket{s}$ & Initial Phase/$\pi$ & \multicolumn{2}{c}{Phase after $\omega_1$ Pulse/$\pi$} && \multicolumn{2}{c}{Phase after $\omega_2$ Pulse/$\pi$} && \multicolumn{2}{c}{ Phase after $\omega_3$ Pulse/$\pi$} & Target Phase/$\pi$\\ 
    \cline{3-4}\cline{6-7}\cline{9-10}\raisebox{12pt}

    + \emph{Perms}& $\phi_s$ & $\phi_s$ & $\phi_s$ (mod~$2\pi$) && $\phi_s$ & $\phi_s$ (mod~$2\pi$) && $\phi_s$ & $\phi_s$ (mod~$2\pi$) & $\phi_s$ \\
    \hline\hline\raisebox{10pt}

    $\ket{0}$$\ket{00000}$ & 0 & 3.125 & 1.125 && 3.125 & 1.125  && 3.125 & 1.125 & 0 \\
    $\ket{0}$$\ket{00001}$ & 0 & 1.125 & 1.125 && 2.125 & 0.125  && 3.125 & 1.125 & 0 \\
    $\ket{0}$$\ket{00011}$ & 0 & 0.125 & 0.125 && 2.743 & 0.743  && 3.125 & 1.125 & 0 \\
    $\ket{0}$$\ket{00101}$ & 0 & 0.125 & 0.125 && 0.507 & 0.507  && 3.125 & 1.125 & 0 \\
    $\ket{0}$$\ket{00111}$ & 0 & 0.125 & 0.125 && 2.743 & 0.743  && 3.125 & 1.125 & 0 \\
    $\ket{0}$$\ket{01011}$ & 0 & 0.125 & 0.125 && 0.507 & 0.507  && 3.125 & 1.125 & 0 \\
    $\ket{0}$$\ket{01111}$ & 0 & 1.125 & 1.125 && 2.125 & 0.125  && 3.125 & 1.125 & 0 \\
    $\ket{0}$$\ket{11111}$ & 0 & 3.125 & 1.125 && 3.125 & 1.125  && 3.125 & 1.125 & 0 \\
    $\ket{1}$$\ket{00000}$ & 0 & 1.125 & 1.125 && 1.125 & 1.125  && 1.125 & 1.125 & 0 \\
    $\ket{1}$$\ket{00001}$ & 0 & 0.125 & 0.125 && 1.125 & 1.125  && 2.125 & 0.125 & 1 \\
    $\ket{1}$$\ket{00011}$ & 0 & 0.125 & 0.125 && 2.743 & 0.743  && 3.125 & 1.125 & 0 \\
    $\ket{1}$$\ket{00101}$ & 0 & 0.125 & 0.125 && 0.507 & 0.507  && 3.125 & 1.125 & 0 \\
    $\ket{1}$$\ket{00111}$ & 0 & 1.125 & 1.125 && 3.743 & 1.743  && 4.125 & 0.125 & 1 \\
    $\ket{1}$$\ket{01011}$ & 0 & 1.125 & 1.125 && 1.507 & 1.507  && 4.125 & 0.125 & 1 \\
    $\ket{1}$$\ket{01111}$ & 0 & 3.125 & 1.125 && 4.125 & 0.125  && 5.125 & 1.125 & 0 \\
    $\ket{1}$$\ket{11111}$ & 0 & 6.125 & 0.125 && 6.125 & 0.125  && 6.125 & 0.125 & 1 \\
    \hline\hline
\raisebox{12pt}

    \textbf{Fidelity}&\textbf{0.5}&\multicolumn{2}{c}{\textbf{0.25}}&&\multicolumn{2}{c}{\textbf{0.634}}&&\multicolumn{2}{c}{\textbf{1}}&\textbf{-}\\
    
    \hline\hline
\end{tabularx}
\caption{Phases $\phi_s$ acquired by basis states $\ket{s}$ after each stage of the composite entangling pulse. Note that the final state is equivalent to the target up to a global phase of $\epsilon=1.125\pi$. The fidelity with the target state at each stage is shown on the final row.}
\label{table:phases}
\end{table*}

\end{document}